\documentclass{article}
\usepackage{spconf,amsmath,epsfig}
\usepackage{amssymb}
\usepackage{array}
\usepackage{multirow}

\newcommand{\real}{{\mathbb R}}
\newcommand{\q}{{\mathrm q}}
\newcommand{\resi}{{\mathrm r}}
\newcommand{\qc}{{\mathrm q}_{\mathrm c}}
\newcommand{\qr}{{\mathrm q}_{\mathrm r}}
\newcommand{\dc}{d_{\mathrm c}}

\newcommand{\Cr}{{\mathcal C}_{\mathrm r}}

\title{Searching in one billion vectors: re-rank with source coding}

\name{
\begin{minipage}{3.5cm}
\begin{center}
Herv\'e J\'egou \\
{\rm {INRIA Rennes}}
\end{center}
\end{minipage}
\hfill
\begin{minipage}{5.0cm}
\begin{center}
Romain Tavenard \\
{\rm {University Rennes I}}
\end{center}
\end{minipage}
\hfill
\begin{minipage}{4.5cm}
\begin{center}
Matthijs Douze\\
{\rm {INRIA Grenoble}}
\end{center}
\end{minipage}
\hfill
\begin{minipage}{4.5cm}
\begin{center}
Laurent Amsaleg \\
{\rm {CNRS, IRISA}}
\end{center}
\end{minipage}
}

\address{~\vspace{-0.8cm}}

\begin{document}

\ninept
\maketitle
\begin{abstract}
Recent indexing techniques inspired by source coding have been shown successful to index billions of high-dimensional vectors in memory. In this paper, we propose an approach that re-ranks the neighbor hypotheses obtained by these compressed-domain indexing methods. In contrast to the usual post-verification scheme, which performs exact distance calculation on the short-list of hypotheses, the estimated distances are refined based on short quantization codes, to avoid reading the full vectors from disk. 

We have released a new public dataset of one billion 128-dimensional vectors and proposed an experimental setup to evaluate high dimensional indexing algorithms on a realistic scale. Experiments show that our method accurately and efficiently re-ranks the neighbor hypotheses using little memory compared to the full vectors representation.
\end{abstract}

\section{Introduction}

Approximate nearest neighbors (ANN) search
methods~\cite{DIIM04,ML09,TFW08,WTF09} are required to handle
large databases, especially for computer vision~\cite{SDI06}
and music retrieval~\cite{CVGLRS08} applications.  
One of the most popular techniques is Euclidean
Locality-Sensitive Hashing~\cite{DIIM04}. 
However, most of these approaches are memory consuming, since several hash
tables or trees are required.  
The methods of~\cite{JDS10a,WTF09}, which embeds
the vector into a binary space, better satisfies the memory
constraint. They are, however, significantly outperformed in terms of the
trade-off between memory usage and accuracy by recent methods that 
cast high dimensional indexing to a source coding problem~\cite{SJ10,JDS11,JDSP10}, 
in particular the product quantization-based method of~\cite{JDS11} 
exhibits impressive results for large scale image search~\cite{JDSP10}. 

State-of-the-art approaches usually perform a re-ranking stage to
produce a ranked list of nearest neighbors. This is always done in
partitioning based method such as~LSH~\cite{DIIM04} or
FLANN~\cite{ML09}, as the neighbor hypotheses are not ranked on output
of the index.  But as shown in~\cite{JDS11}, this post verification
is also important for methods based on binary~\cite{JDS10a,WTF09}
or quantized codes~\cite{SJ10,JDS11,JDSP10}, as the ranking provided on 
output of the large scale search is significantly improved 
when verifying the few first hypotheses. 
In all these approaches, re-ranking is performed 
by computing the exact Euclidean distance between each query
and the returned hypotheses. 
For large datasets, this raises a memory issue: 
the vectors can not be stored in central memory and must 
be read from disk on-the-fly. 
Moreover, the vectors are to a large extent accessed randomly, 
which in practice limits the number 
of hypotheses that can be verified. 

In this paper, we propose a new post-verification scheme 
in the context of compression-based indexing methods. 
We focus on the method presented in~\cite{JDS11},
which offers state-of-the-art performance, outperforming 
the FLANN which was previously shown to outperform LSH~\cite{ML09}. 
It also provides an explicit approximation of the indexed vectors. 
Our algorithm exploits the
approximation resulting from the first ranking, and refines 
it using codes stored in RAM.
There is an analogy between this approach and 
the scalable compression techniques proposed in the last decade~\cite{T00}, 
where the term ``scalable'' means that
a reconstruction of the compressed signal is refined 
in an incremental manner by successive description layers. 

In order to evaluate our approach, we introduce a
dataset of one billion vectors extracted from millions of images 
using the standard SIFT descriptor~\cite{Low04}. 
Testing on a large scale is important, as most ANN methods are usually
evaluated on sets of unrealistic size, thereby ignoring memory issues 
that arise in real applications, where billions of vectors have to be handled~\cite{JDS10a,MPCM10}. 
The groundtruth nearest-neighbors have been computed for 
10000 queries using exact distance computations. 
To our knowledge, this set is the largest ever released to evaluate 
ANN algorithms against an exact linear scan on real data:
the largest other experiment we are aware of is the one reported in~\cite{amsaleg09c}, 
where a private set of 179 million vectors was considered. 
\cite{CPSK10} reports an experiment on 1 billion vectors, 
but on synthetic data with a known model exploited by the algorithm. 

Experiments performed on this dataset show
that the proposed approach offers an alternative to 
the standard post-verification scheme. The precision of the search 
is significantly improved by the re-ranking step, 
leading to state-of-the-art performance on this scale, without accessing the disk.

\section{Context: compression based indexing}
\label{sec:sourecoding}

In this section, we briefly review the indexing method
of~\cite{JDS11}, which finds the approximate $k$ nearest neighbors
using a source coding approach. For the sake of presentation, 
we describe only the Asymmetric Distance Computation (ADC) method 
proposed in~\cite{JDS11}. 
We also assume that we search for the nearest neighbor (i.e., $k=1$), 

\medskip

Let $x \in \real^d$ be a query vector and ${\mathcal Y}=\{y_1,\dotsc,y_n\}$  
a set of vectors in which we want to find the nearest neighbor NN$(x)$ of $x$. 
The ADC approach consists in encoding each vector $y_i$ 
by a quantized version $c_i=\qc(y_i) \in \real^d$. 
For a quantizer $\qc(.)$ with $K$ centroids, 
the vector is encoded by $b_{\mathrm{c}}=\log_2(K)$ bits, assuming $K$ is a power of 2. 
An approximate distance $\dc(x,y_i)$ between a query $x$ and a database vector 
is computed as 
\begin{equation}
\dc(x,y_i)^2= \|x - \qc(y_i)\|^2. 
\label{equ:adc}
\end{equation}

The approximate nearest neighbor~NN$_a(x)$ of~$x$ is obtained by 
minimizing this distance estimator:
\begin{equation}
\text{NN}_a(x) = \arg\min_i \dc(x,y_i)^2= \arg\min_i \|x - \qc(y_i)\|^2, 
\label{equ:searchann}
\end{equation}
which is an approximation of the exact distance calculation
\begin{equation}
\text{NN}(x) = \arg\min_i \|x - y_i\|^2. 
\label{equ:searchnn}
\end{equation}
Note that, in contrast with the binary embedding method of~\cite{WTF09}, 
the query $x$ is not converted to a code: there is no approximation
error on the query side. 

To get a good vector approximation, $K$ should be large ($K=2^{64}$
for a 64 bit code). For such large values of $K$, learning a $K$-means
codebook is not tractable, neither is the assignment of the vectors to their nearest centroids.  
To address this issue, \cite{JDS11} uses a product quantizer, 
for which there is no need to explicitly enumerate the centroids.
A vector $y \in \real^d$ is first split into $m$ subvectors $y^1, ..., y^m \in \real^{d/m}$. A product quantizer is then defined as a function 
\begin{equation}
\qc(y) = \big(\q^1(y^1),...,\q^m(y^m)\big), 
\end{equation}
which maps the input vector $y$ to a tuple of indices by separately quantizing 
the subvectors. Each individual quantizer $\q^j(.)$ has $K_{\mathrm{s}}$ reproduction values, 
learned by $K$-means. To limit the assignment 
complexity, ${\mathcal O(m \times K_{\mathrm{s}})}$, 
$K_{\mathrm{s}}$ is set to a small value (e.g. $K_{\mathrm{s}}$=256). 
However, the set of $K$ centroids induced by the product quantizer $\qc(.)$
is large, as $K=(K_{\mathrm{s}})^m$. 
The squared distances in Equation~\ref{equ:searchann} are computed 
using the decomposition 
\begin{equation}
\dc(x,y)^2=\|x - \qc(y)\|^2 = \sum_{j=1,...,m} \|x^j - \q^j(y^j)\|^2, 
\label{equ:dc}
\end{equation}
where $y^j$ is the $j$\textsuperscript{th} subvector of $y$. 
The squared distances in the sum are read from look-up tables. 
These tables are constructed on-the-fly for a given query, 
prior to the search in the set of quantization codes, from each 
subvector $x^j$ and the $k_{\mathrm{s}}$ centroids associated with 
the corresponding quantizer $\q^j$. 
The complexity of the table generation is ${\mathcal O(d \times K_{\mathrm{s}})}$. 
When $K_{\mathrm{s}} \ll n$, this complexity is negligible compared to
the summation cost of ${\mathcal O}(d \times n)$ in Equation~\ref{equ:searchann}. 
\medskip

This approximate nearest neighbor method implicitly 
sees multi-dimensional indexing as a vector approximation problem: 
a database vector $y$ can be decomposed as  
\begin{equation}
y = \qc(y) + \resi(y), 
\label{equ:approx}
\end{equation}
where $\qc(y)$ is the centroid associated with $y$ 
and $\resi(y)$ the error vector resulting from the quantization, 
called the \emph{residual} vector. 
It is proved~\cite{JDS11} that the 
square error between the distance and its estimation 
is bounded, on average, by the quantization error. 
This ensures, asymptotically, perfect search results 
when increasing the number of bits allocated for the quantization indexes.
\medskip

The ADC indexing method is fully parametrized by the number of subvectors $m$ 
and the total number of bits $b_{\mathrm c}$. 
In the following, we set~$b_{\mathrm c}=8$ (i.e., $K_{\mathrm s}=256$), 
as suggested in~\cite{JDS11}, which means that we use exactly $m$ bytes 
per indexed vector.

\section{Re-ranking neighbors using source coding}
\label{sec:method}

\subsection{Refinement: principle}

The objective of the method proposed in this paper is to avoid 
the costly post-verification scheme adopted in most 
state-of-the-art approximate search techniques~\cite{DIIM04,ML09}.
The idea is to take advantage of the information on the
database point provided by the 
indexing. This is possible when using the ADC method~\cite{JDS11}: 
this search algorithm provides an explicit approximation $\qc(y)$ 
of database vector $y$. 

We first assume that the first retrieval stage returns a set of $k'$ 
hypotheses. These vectors are the one for which a post-verification 
is required. For each database vectors~$y_i$, the error vector is equal to
\begin{equation}
\resi(y) = y - \qc(y). 
\end{equation}

The proposed method consists in reducing the energy of this residual vector 
to limit the impact of the approximation error on the estimated distances. 
This is done by encoding the residual vector $\resi(y)$ 
using another product quantizer $\qr$ defined by its 
reproduction values~$\Cr$:
\begin{equation}
\qr(\resi(y)) = \arg \min_{c \in \Cr} \|\resi(y)-c\|^2,
\end{equation}
where the product quantizer~$\qr(.)$ is learned on an independent 
set of residual vectors. Similar to $\qc(.)$, the set of reproduction 
values~$\Cr$ is never exhaustively listed, as all operations are 
performed using the product space decomposition. 

The coded residual vector can be seen as the ``least significant bits'', 
except that the term ``bits'' 
usually refers to scalar quantization. 
An improved estimation~$\hat{y}$ 
of~$y$ is the sum of the approximation vector and 
the decoded residual vector:
\begin{equation}
\hat{y} = \qc(y) + \qr(\resi(y)).  
\label{equ:hatyi}
\end{equation}

\begin{figure}
\includegraphics[width=0.95\linewidth]{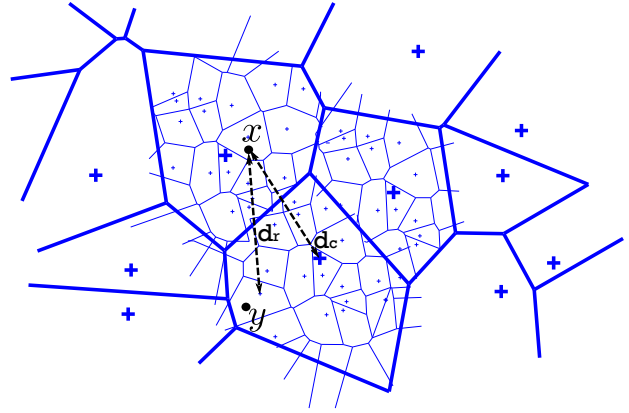}
\caption{ Illustration of the proposed refinement process. For each
  database vector $y$, the distance $d_\mathrm{c}(x,y)=d(x,\qc(y))$ is computed to 
  build the short-list of potential nearest neighbors.  For selected
  $y$ vectors, the distance is re-estimated by $d_\mathrm{r}(x,y)$, which
  is obtained by computing the distance between $y$ and its improved
  approximation $d_\mathrm{r}=\qc(y)+\qr(y-\qc(y))$.}
\vspace{-0.5cm}
\label{fig:incremental}
\end{figure}

As shown in Figure~\ref{fig:incremental}, this estimator 
will be used at search time to update the distance estimation between 
the query $x$ and the database vectors $y$ that are selected as 
potential neighbors:
\begin{equation}
d(x,y)^2 \approx d_\mathrm{r}(x,y)^2 = \|\qc(y) + \qr(\resi(y)) - x\|^2. 
\end{equation}

The refinement product quantizer $\qr$ is parametrized 
by its number of subquantizers and the total number of bits. 
Similar to $\qc$, we use 8 bits per subquantizer. 
Therefore the only parameter for the refinement quantizer 
is the number~$m'$ of bytes for each code. 
The total memory usage per indexed vector is $m+m'$ bytes. 

\subsection{Algorithm}

This subsection details how the refinement codes are used 
to re-rank the hypotheses provided by the ADC. The resulting 
approach will be denoted by ADC+R in the experimental section. 
As for most indexing algorithms, we distinguish between the offline stage 
and the query stage, during which the system needs to be very efficient. 
The offline stage consists in learning the indexing parameters and 
building the indexing structure associated with a vector dataset. 
It is performed as follows. 
\begin{enumerate}
\item The quantizers $\qc(.)$ and $\qr(.)$ are learned on a 
training set. 
\item The vector dataset ${\mathcal Y}=\{y_1,\dotsc,y_n\}$ to be indexed 
is encoded using~$\qc$, producing codes $\qc(y_i)$ for $i=1,\dotsc,n$. 
\item The residual vectors are encoded, producing 
the codes $\qr(y_i-\qc(y_i))$ associated with all the indexed vectors.
\end{enumerate}
\medskip

\noindent Searching a query vector~$x$ proceeds as follows:
\begin{enumerate}
\item The ADC distance estimation is used to generate a list ${\mathcal L}$ of $k'$ 
hypotheses. The selected vectors minimize the 
estimator of Equation~\ref{equ:dc}, which is computed directly 
in the compressed domain~\cite{JDS11}. 
\item For each vector $y_i \in {\mathcal L}$, 
the approximate vector~$\hat{y}_i$ is explicitly reconstructed using 
the first approximation~$\qc(y_i)$ and the coded residual 
vector $\qr(y_i)$, see Equation~\ref{equ:hatyi}. 
The squared distance estimator~$d(x,\hat{y}_i)^2$ is subsequently computed. 
\item The vectors of ${\mathcal L}$ associated with the $k$ smallest 
refined distances are computed. 
\end{enumerate}


On output, we obtain a re-ranked list of $k$ approximate nearest neighbors. 
The choice of the number~$k'$ of vectors in the short-list depends 
on parameters $m$, $m'$, $k'$ and on the distribution of the vectors. 
In order for the post-verification scheme to have a negligible complexity, 
we typically set the ratio $k'/k$ to 2. 

\subsection{Non exhaustive variant}

Up to now, we have only considered the ADC method of~\cite{JDS11}, 
which requires an exhaustive scan of the dataset codes. 
Note however that the re-ranking method proposed in this paper 
can be applied to any method for which an approximate reconstruction of 
the indexed vectors is possible, e.g.,~\cite{SJ10}. 
In particular, in the experimental section 
we evaluate our approach with the IVFADC variant of~\cite{JDS11}, 
that avoids the aforementioned exhaustive scan by using an inverted file structure. 
This requires an additional coarse quantizer.


Adapting our re-ranking method to the IVFADC method 
is straightforward, as this method also provides 
an explicit approximation of the indexed vectors. 
In addition to the numbers~$m$ and~$m'$
of bytes used to encode the vector, 
this variant requires two additional parameters: the number~$c$ 
of reproduction values of the coarse quantizer and the number~$v$ 
of inverted lists 
that are visited for a given query. 
The main advantage of this variant is to compute the distance 
estimators only for a fraction (in the order of $v/c$) 
of the database, at the risk 
of missing some nearest neighbors if $v/c$ is not large enough. 
Note finally that the memory usage is increased by $\log_2(c)$ bits 
(typically 4 bytes), due to the inverted file structure. 
In the following, the IVFADC method used jointly with our 
re-ranking method will be denoted by IVFADC+R. 

\section{Experiments}
\label{sec:experiments}

\subsection{BIGANN:  a billion-sized evaluation dataset}
\label{sec:bigann}

To evaluate ANN search methods, we propose a new 
evaluation dataset available online:
\texttt{http://corpus-texmex.irisa.fr}. 
This benchmark, called BIGANN, consists of 128-dimensional SIFT 
descriptors (widely adopted image descriptors~\cite{Low04}) extracted 
from approximately 1 million images. 
It comprises three distinct subsets:
\begin{itemize}
\item base vectors: one billion vectors to search in 
\item query vectors: 10000 vectors that are submitted to the system
\item learning vectors: a set of 100 million vectors to compute
 the parameters involved in the indexing method. 
\end{itemize}

The groundtruth has been pre-computed: for each query, 
we provide the $k$ nearest neighbors that are obtained 
when computing exact Euclidean distance, 
as well as their square distance to the query vector. 
The groundtruth for smaller sets ($n$=1M, 2M, 5M, 10M, ..., 200M vectors) 
is also provided. 
As our own approach does not require many training vectors, we only 
used the first million vectors from the learning set. 
All measurements (accuracy and timings) were averaged over the 1000 
first queries. 

\subsection{Evaluation protocol}

The search quality is measured by the recall@$r$ measure, i.e., the proportion
of queries whose nearest neighbor is ranked in the first $r$ positions. 
The curve obtained by varying $r$ corresponds to the
distribution function of the ranks, and the point $r$=1 
corresponds\footnote{In practice, we are often interested in retrieving the $k$ nearest
neighbors ($k>1$) and not only  the nearest neighbor.  We do not
include these measures in the paper, as qualitative conclusions 
for $k$=1 remain valid for $k>1$.}
to the ``precision'' measure used in~\cite{ML09} to evaluate ANN methods.
Also, the recall@$r$ is the fraction of queries for which
the nearest neighbor would be retrieved correctly if a short-list of $k=r$
vectors was verified using exact Euclidean distances. 

The efficiency is measured by actual timings. 


\subsection{Evaluation of the proposed approach}

Unless explicitly specified, our approach is evaluated 
by querying in the whole BIGANN set (i.e., one billion vectors). 
The performance of the ADC and IVFADC algorithms are given for reference, 
and compared with the re-ranked versions ADC+R and IVFADC+R. 
In all experiments, we have set $k$=10000 and $k'$=20000. 
In addition, for the IVFADC+R we have fixed $c$=8192 and $v$=64, 
which means that the query is compared to approximately 
1/128\textsuperscript{th} of the indexed vectors. 
\medskip

\begin{figure}[t]
\centering
\includegraphics{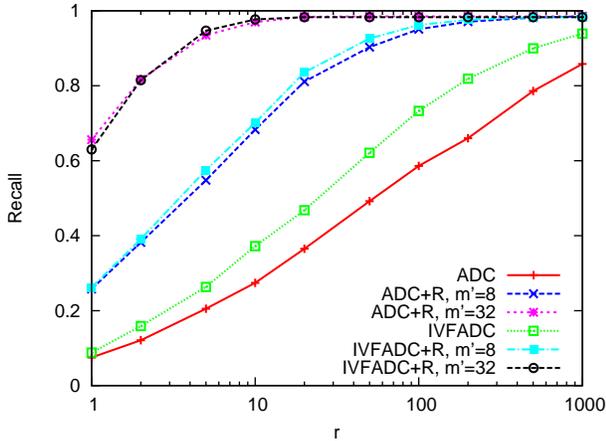}
\vspace{-0.2cm}
\caption{Searching in one billion vectors: 
impact of the re-ranking stage on the search accuracy (recall@r).
$m$=8.
\label{fig:pqr_vs_pq}}
\vspace{-0.2cm}
\end{figure}

{\noindent \bf The re-ranking gain:} We first consider the improvement 
brought by the re-ranking stage compared with the reference methods.
Figure~\ref{fig:pqr_vs_pq} shows the importance of this re-ranking stage 
on the recall@r measure: the performance of PQ+R (resp. IVFPQ+R) 
is significantly better than that of ADC (resp. IVFADC).

These observations are confirmed by Table~\ref{tab:m_m'_user_times}, 
which additionally shows that the impact of the re-ranking stage 
on the query time is limited. As already reported in~\cite{JDS11}, 
the IVFADC version is better than the ADC method, 
at the cost of 4 additional bytes per indexed vector. 
As expected, this observation remains 
valid when comparing IVFADC+R with ADC+R. 
\medskip

\begin{table}[t]
{\small
\centering
\begin{tabular}{cccccc}
\hline
Method	& $m'$	& recall@1	& @10	& @100	& time/query \\
\hline


ADC     & 0	& 0.075	& 0.274	& 0.586	& 5.626 \\
\hline
\multirow{3}{*}{ADC+R}	
        & 8	& 0.258	& 0.683	& 0.951	& 5.686 \\
		& 16	& 0.434	& 0.895	& 0.982	& 5.692 \\
		& 32	& 0.656	& 0.970	& 0.985	& 5.689 \\
\hline
IVFADC  & 0	& 0.088	& 0.372	& 0.733	& 0.074 \\
\hline
\multirow{3}{*}{IVFADC+R}  
	& 8	& 0.262	& 0.701	& 0.962	& 0.116 \\
	& 16	& 0.429	& 0.894	& 0.982	& 0.119 \\
	& 32	& 0.630	& 0.977	& 0.983	& 0.120 \\
\hline
\end{tabular}
\caption{Performance and efficiency measured on 1 billion vectors, $m$=8. 
The query time is measured in seconds per query. 
The timings validate the limited impact of the re-ranking stage on efficiency. 
\label{tab:m_m'_user_times}
}}
\end{table}

{\noindent \bf Performance for a fixed memory usage:} 
Table~\ref{tab:m_m'_mem} shows that, for a given memory usage (i.e., for $m+m'$ constant),
the re-ranking approach ADC+R achieves similar recall as ADC, 
at a lower computing cost. The search is approximately two times faster, 
as the search time is dominated by the first retrieval stage, whose complexity 
is asymptotically linear in $m$ and $n$ for large values of $n$. 
One can also observe that near perfect neighbors are obtained by 
increasing the number~$m'$ of bytes used to re-rank the queries. 
\medskip

\begin{table}[t]
{\small
\centering
\begin{tabular}{cccccl}
\hline
bytes/vector  & $m$	& $m'$	& recall@1	& @10	& @100  \\
\hline

\multirow{2}{*}{8}
& 8	& 0	& 0.075	& 0.274	& 0.586	\\
& 4	& 4	& 0.056	& 0.223	& 0.504	\\
\hline
\multirow{2}{*}{16}
& 16	& 0	& 0.245	& 0.671	& 0.952	\\
& 8	& 8	& 0.258	& 0.683	& 0.951	\\
\hline
\multirow{2}{*}{32}
& 32	& 0	& 0.487	& 0.956	& 0.999	 \\
& 16	& 16	& 0.571	& 0.977	& 1.000	 \\
\hline
\multirow{2}{*}{64}
& 64	& 0	& 0.791	& 1.000	& 1.000	 \\
& 32	& 32	& 0.832	& 0.999	& 1.000	 \\

\hline
\end{tabular}
\vspace{-0.2cm}
\caption{Comparative accuracy of ADC ($m'$=0) and ADC+R for the same \emph{total} amount 
of memory (bytes per indexed vector). In these experiments, the ADC+R approach with $m=m'$ 
is approximately 2$\times$ more efficient that the ADC approach, as 
the cost of the the re-ranking stage is almost negligible, as shown Table~\ref{tab:m_m'_user_times}. 
These experiments have been performed on the whole BIGANN set. 
\label{tab:m_m'_mem}
}
}
\vspace{-0.2cm}
\end{table}

\begin{figure}[t]
\centering
\includegraphics{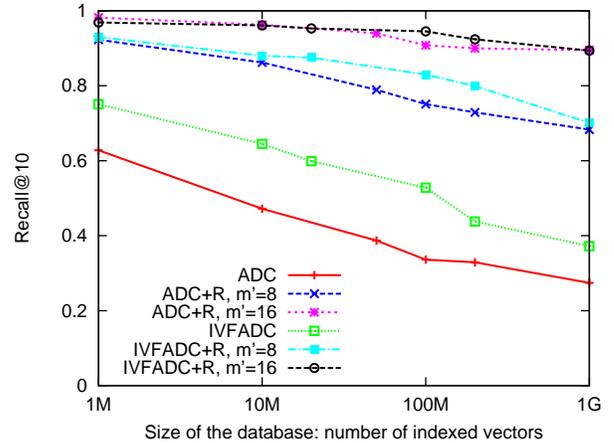}
\vspace{-0.2cm}
\caption{Impact of the database size on search performance measured 
by recall@10. $m$=8.\label{fig:rec_vs_nbase}}
\vspace{-0.6cm}
\end{figure}

{\noindent \bf Impact of the dataset size:}
Figure~\ref{fig:rec_vs_nbase} shows the impact of the database size
on the recall@10 measure. The re-ranking stage becomes 
more important as the database grows in size, due to an increasing 
number of outliers. Interestingly, the quality of the search 
degrades more gracefully when using a precise 
post-verification scheme (with $m'$=16 bytes). 


\section{Conclusion}

In this paper, following recent works on multi-dimensional indexing 
based on source coding, 
we propose a method to re-rank the vectors with a 
limited amount of memory, thereby avoiding costly disk accesses. 
Refining the neighbor hypotheses strongly improves recall 
at a rather low cost on response time.
Our experimental validation is performed on a new public dataset 
of one billion vectors. 

\section{Acknowledgements}

This work was partly realized as part of the Quaero Programme, funded by OSEO, French State agency for innovation.

\end{document}